**Climate control in termite mounds: A review and new insights from X-ray tomography and flow field simulations**

Nengi F Karibi-Botoye [1*], Guy Theraulaz [2], Bagus Muljadi [3], Vasily Demyanov [1] and Kamaljit Singh [1]

[1] *Institute of GeoEnergy Engineering, School of Energy, Geoscience, Infrastructure & Society, Heriot-Watt University, EH14 4AS Edinburgh, United Kingdom.*
[2] *Centre de Recherches sur la Cognition Animale (CRCA), Centre de Biologie Intégrative (CBI), Université de Toulouse,CNRS, UPS, Toulouse, France.*
[3] *Department of Chemical and Environmental Engineering, University of Nottingham, NG7 2RD Nottingham, United Kingdom.*

**Abstract**

Investigating the thermoregulation and ventilation processes in termite mound holds great interest, in particular for its potential implications in advancing eco-friendly building designs. In this article we discuss major results on these processes in termite mounds of varying sizes and ventilation types. Additionally, we propose the integration of X-ray tomography to gain insights into the contribution of architectural features of termite mounds to thermoregulation and ventilation. Finally, we assess the contributions of numerical flow field simulations to this research domain. Our objective is to consolidate existing knowledge, identify research gaps, and propose an interdisciplinary approach to foster our understanding of temperature regulation and gas exchange in termite mounds.

Author for correspondence (E-mail: nfk3@hw.ac.uk).

# Contents

I.  Introduction

Termites have been considered as supreme builders within the animal kingdom, skillfully combining form and function in their structures (Claggett *et al.,* 2018). Their mounds have been a source of interest for biologists, engineers, and architects due to their unique ability to manage heat, humidity, and gases without the use of mechanical devices despite changes in external climatic conditions (Korb, 2003; King, Ocko, & Mahadevan, 2015). The understanding and the potential replication in human buildings of the processes underlying temperature regulation and ventilation in termite mounds could lead to advancement in the fields of architecture and engineering, enabling buildings to meet the needs of occupants while reducing energy requirements and carbondioxide ($CO_2$) emissions (Turner & Soar, 2008; Ball, 2010; Yuan *et al*., 2017; Pawlyn, 2019; Gorb & Gorb, 2020).

Termite mounds vary in size and shape all of which can impact the thermoregulatory and ventilation processes. Their size could range from a few centimeters to several meters with colony size ranging from hundreds to millions (Nicholas M. C.; unpublished data; Korb, 2003; Ocko, Heyde & Mahadevan, 2019; Räsänen *et al.,* 2023). The structural integrity of these mounds can persist for several decades to centuries (Martin *et al.,* 2018; Zachariah *et al.,* 2020a; Zachariah *et al*., 2020b). The mound skin or walls serves as the boundary layer between the mound internals and the external environment, and it contains a network of micropores of varying sizes (Abou-Houly Eid H. unpublished data; Singh *et al.* 2019). The internal structure of the mound typically consists of a complex network of tunnels, chimneys, surface conduits, royal cells, lateral conduits, ridges, turrets, which vary based on the species of termites. Mounds have been described as consisting of four distinct zones: the outer mantle, the transition zone, the accumulation zone, and the nest (Erens *et al.,* 2015).

Numerous studies on mound formation have indicated that their construction mainly involves self-organized processes resulting from local stigmergic interactions that are coupled with template effects resulting from temperature and/or gas gradients (Bonabeau *et al.,* 1997; Theraulaz, Bonabeau, & Deneubourg, 1998; Theraulaz and Bonabeau, 1999; Ocko *et al.,* 2019; Perna and Theraulaz, 2019, Heyde *et al.,* 2021).

Termite mounds can be epigeal – i.e., built above the soil surface, subterraneous – when the whole nest or just a part of it is built below the ground, and arboreal – when nest is built on trees (Grassé, 1984). The term "mound" is mostly used interchangeably with "nest". In this article we use the term *mound* to describe the large epigeal part, whereas the interior of the mound where the fungus (if present) and termite queen reside is referred to as *nest*.

The termites living in mounds typically measure around 3 to 5 millimeters in height and 4 to 14 millimeters in length. They consist of three main castes – the reproductive (queen and king), the soldier and the worker castes (Bignell, Roisin & Lo, 2010; Mariod, Saeed & Hussein, 2017). The construction of termite mounds is commonly carried out by the worker caste, employing a mixture of materials such as saliva, wood, excrement, and soil (Schmidt, 1955; Noirot & Darlington, 2000; Zachariah *et al*., 2020a; Jost, 2021).

Termites inhabit every continent except Antarctica, showcasing their presence in a wide range of climates and their capacity to adapt (Claggett *et al.,* 2018). These termites could either be fungus farming or non-fungus farming feeding on soil, grass, leaves and wood. Fungus farming termites usually produce higher $CO_2$ and require specific temperature conditions (Aanen & Eggleton, 2005).

Broadly speaking, mounds can have open or closed ventilation types. Open ventilation happens in mounds that have visible millimetre (mm) scale holes in the walls, whereas closed ventilation happens when there are no visible holes in the mound walls and thus ventilation is achieved via microscale pores in the outer walls (Korb & Linsenmair, 2000; Singh *et al.*, 2019). Some species, such as *Odontermes obesus*, *Procornitermes araujoi*, *Trinervitermes geminatus*, *Macrotermes michaelseni* and *Macrotermes bellicosus* (Jost, 2021), have been documented to exhibit closed ventilation. Whereas, *Odontotermes transvaalensis*, *Odontotermes tanganicus*, and *Macrotermes subhyalinus* have been reported to have open ventilation (Darlington, 1985).

Despite recognizing the importance of thermoregulatory and ventilation processes in termite mounds, a comprehensive understanding of the underlying mechanisms remains elusive. Our limited understanding can be attributed to the challenges of directly measuring conditions within the mound, particularly in relation to ventilation. Additionally, there is a lack of complete visualization of the internal mound structure and how various architectural and structural characteristics within the mound impact its thermal and flow properties. Furthermore, there has been a limited amount of modelling conducted in this research field. The difficulty of understanding the mechanisms that control temperature and ventilation is even greater if one consider the diversity of termites with 2951 species and 297 genera described (Constantino, 2018), each of them having its own mound design. Therefore, it is necessary to further investigate existing concepts, identify gaps in knowledge, and discuss future research approaches required to facilitate a comprehensive understanding of these topics.

In this article, we first examine the relationship between mound temperature and factors like the mound size, the ventilation types, the occupancy, the ambient temperature, and the soil temperature. We then discuss the current understanding of ventilation and humidity controls within the mound. We also examine the role of X-ray tomography in understanding climate control in mounds, its past applications, and its potential integration with numerical simulations in the future. Finally, we discuss field works and numerical simulation carried out in the mound. To conclude, we suggest how the X-ray tomography and numerical simulation techniques could address knowledge gaps in the future.

## II. Thermoregulation in termite nest

Temperature regulation is crucial for termites as they require an environment with a consistently controlled temperature and high humidity as well as an efficient system for exchanging gases with their surroundings. Temperature regulation is especially important for termites that grow fungi, as the fungi require specific temperature ranges to grow (Aanen & Eggleton, 2005; Aiki, Pirk, & Yusuf, 2019). Termites regulate temperature within their nests through both passive and active methods. Active regulation involves termites participating in temperature control, such as clustering to generate metabolic heat (Jones & Oldroyd, 2006). Passive regulation occurs when temperature control is achieved indirectly through the termites' selection of a suitable mound site, such as under a shade or in the open air, or through the design of the mound structure and material used to build the mound, which can allow or inhibit heat flow and alter the orientation of the mound (Perez & Aron, 2020). Termites primarily rely on passive temperature regulation (Jones & Oldroyd, 2006).

### (1) Temperature profiles within termite mound

Several studies have been carried out to investigate the temperature profiles at different locations within the termite mounds (Turner, 2001; Field & Duncan, 2013; King *et al.,* 2015; Ocko *et al.,* 2017; Vesala *et al.,* 2019). The temperature profiles provide information about temperature stability within the mounds and how it varies from location to location within the mound. Temperature is usually measured in different parts of the termite mound, including the surface channels (channels closer to the outer walls), the internal channels deeper into the mound, and within the nest where the queen resides (Korb & Linsenmair, 2000; Ocko *et al.,* 2017; Vesala *et al.,* 2019). These temperature measurements are usually carried out using ibuttons, which are able to log temperature for a long time, thermistors, temperature sensors and thermometers.

The temperatures within the mound in locations like the surface channels and internal channels have been reported to vary from day to day but this variation is much less compared to the ambient air temperature. This variation in temperature increases from month to month but still much less than the variation in ambient air temperature. On the other hand, the temperature within a nest has been the most stable when compared to other locations in the mound. Its day-to-day variation is minimal. Vesala et al., (2019) reported diurnal temperature changes in the nest between 0 and 4 ºC compared to a difference of up to 20 ºC in ambient air. Korb & Linsenmair (2000) reported approximately 1 ºC change in daily temperature in the nest, and Korb (2003) reported only about 2 ºC variation in temperature annually in the nest while a variation up to 35 ºC was recorded in ambient air. Moreover, within the same season, one can observe very small variations in nest temperature. This variation becomes more pronounced across seasons, from winter to summer, which can be up to 6 ºC difference (J. S. Turner, 2001).

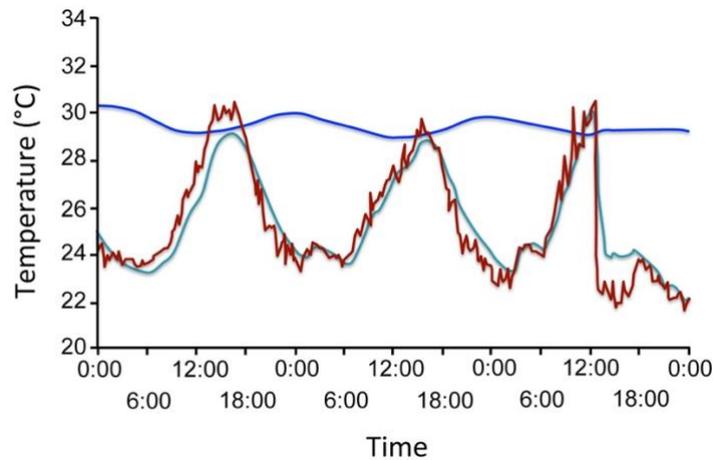

*Figure 1. Temperature profile in the mound and ambient temperature over the course of three days. Temperature of air 1 cm above the surface of the mound in light blue (———), inside the air channels in red (———) and inside the nest in dark blue (———) (adapted from Korb & Linsenmair, 2000).*

Typical temperature profiles at different locations in the termite mound are shown in **Figure 1**. These profiles suggest that there is temperature stability inside termite nests and temperature regulation does indeed occur within the nest because the influence of outside temperature is minimal.

Notably, these profiles are primarily obtained from large and more advanced termite species like the *Macrotermes* species. Understanding how temperature profiles vary in species that build smaller mounds is also important. While there exists a limited number of studies on species building smaller and less developed mound architecture, they suggest an increased temperature variation within the mound, particularly in those with thin walls (Lüscher, 1961). It is important to focus on further studies in this area to pinpoint the factors or structures influencing temperature regulation in smaller termite mounds.

**(2) Temperature control mechanisms in termite mounds**

In this section we examine several temperature controlling mechanisms in different termite mounds.

**(a) Temperatures in inhabited and uninhabited mounds**

The heat within termite mounds comes mainly from two sources, the sun (abiotic heat), and the metabolic activity of termites or the fungus they cultivate within the nest (biotic heat). To assess the impact of biotic heat, one must compare the temperature difference between nests with inhabitants and those without.

In several studies conducted in both inhabited and uninhabited termite mounds, it has been observed that their nest temperature remained consistently stable despite changes in ambient temperature (Holdawayt & Gayt, 1948; T Greves, 1964; Korb & Linsenmair, 2000; Field & Duncan, 2013). However, the nests occupied by termites exhibited higher temperatures compared to the unoccupied ones. For instance, in the nest of *M. bellicosus*, the temperature difference amounted to approximately 3 °C, with inhabited nests measuring around 29.9 °C and uninhabited ones around 26.78 °C. This temperature difference can be attributed to the metabolic heat produced by the termites and fungi. In nest constructed by *Trinervitermes trinervoides* termites, the difference in temperature was about 6.1°C between inhabited and uninhabited nests in the winter months and 3 to 3.8 °C in the summer months.

It is worth noting that the optimal temperature for cultivating fungus is around 30 °C; hence biotic heat raises the nest temperature to the optimal required level. In non-fungus farming termites, there may be no requirement for additional temperature increase.

These observations indicate that temperature regulation occurs regardless of the presence of termites within the nest; however, termites do contribute to elevating the temperature within the nest via biotic heat. Therefore studying the architecture of termite mounds is key to understanding how thermoregulation is achieved.

**(b) Temperatures in mounds of different sizes**

In addition to examining the impact of termite inhabitation on nest temperature, it is crucial to evaluate the effect of mound size on temperature. It has been reported that large mounds are relatively less affected by ambient temperature in comparison to smaller mounds (Josens, 1971; Korb & Linsenmair, 2000; Field & Duncan, 2013; Jost, 2021). Nevertheless, the definition of "large mounds" has varied across these studies. In the case of *M. bellicosus,* large mounds were identified as those exceeding 2 meters in height or colonies that have persisted despite disruptions to their mounds. Conversely, for *T. trinervoides*, large mounds were characterized by heights ranging from 60 to 70 centimetres and circumferences between 400 to 500 centimetres. These large mounds have demonstrated reduced fluctuations in the nest temperature from month to month, contrasting with the variability seen in smaller mounds. Jost (2021) asserted that maintaining nearly constant temperatures seems achievable primarily within large mounds.

Furthermore, the temperature in large colonies of *T. trinervoides* has been reported to be greater than that of smaller colonies, which is 3 °C higher in the winter and 5.4 °C in the summer, potentially due to their ability to produce more biotic heat because of their larger size (Field & Duncan, 2013).

As mentioned previously, uninhabited mounds had a temperature of approximately 27 °C, as opposed to the optimal temperature for fungus growth of 30 °C. If the heat capacity of mounds is between 2500 – 2750 $KJC^{-1}$, raising the temperature by 3 °C requires about 7500 – 8250 KJ. Large colonies containing approximately 2 million termites could supply 8640 KJ/day, while small colonies only supplied 860 KJ/day. As a consequence, in small colonies heat production is not enough to raise the temperature of the mound up to the optimal level, leading to suboptimal temperatures for fungus farming. Suboptimal temperature conditions can be detrimental to the cultivation of fungi in termite nests, as some fungus species are unable to thrive at temperatures below their optimal range usually between 29 – 30 °C (Thomas, 1981). Research on different *Termitomyces* fungus species has revealed that each species has distinct temperature requirements. Notably, one of the cultivated fungal species appears to be relatively intolerant to lower temperatures (Vesala *et al.,* 2019).

The enhanced temperature stability observed in large mounds also arises from their greater thermal mass, because they possess a lower surface area to volume ratio and a higher heat capacity. This characteristic enables them to absorb and retain more heat, gradually releasing it into the external environment, consequently leading to reduced temperature variability. Additionally, in grass-feeding species like *T. trinervoides*, large mounds have a larger periphery area to store grass, potentially serving as a form of insulation for the nest.

**(c) Temperatures in mounds with different ventilation systems**

Beyond differences in size, termite mounds can exhibit distinct ventilation types, depending on whether their architecture is closed or open as shown in **Figure 2**.

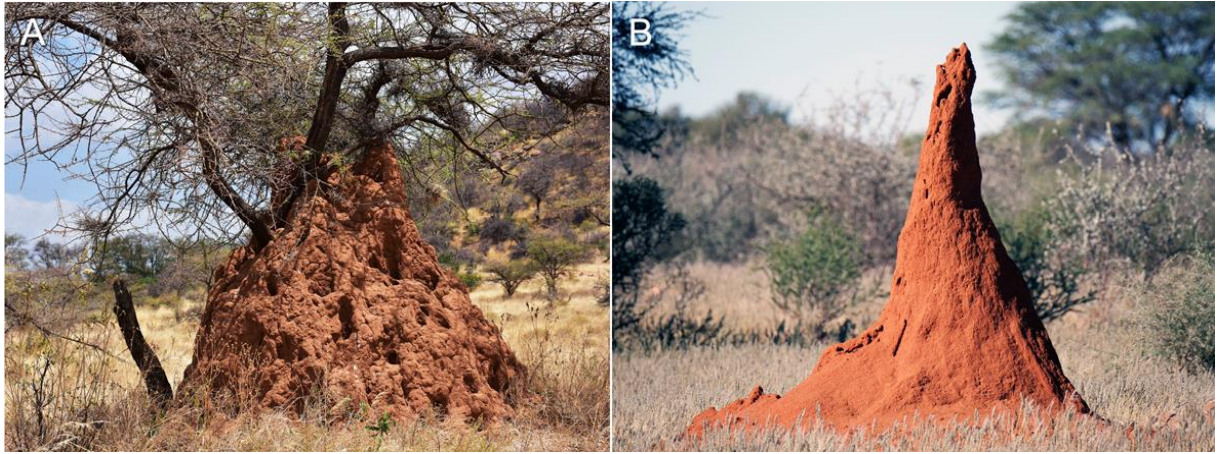

*Figure 2. Termite mound showing different ventilation systems (A) open ventilation, Embu, Kenya, Photograph courtesy of orientalizing (flickr.com) (B) closed ventilation, Hardap, Namibia, photograph by Nevil Rimes.*

The type of ventilation system can influence nest temperatures. For instance in the closed mounds built by *M. michaelseni*, nest temperatures consistently exceeded those found in open mounds of *M. subhyalinus* and showed less temperature fluctuations (Vesala *et al.,* 2019). This difference can potentially be attributed to the absence of wall openings in closed mounds, thereby preventing heat loss and minimizing the influx of air in comparison to nests with open ventilation. Additionally, mounds with open ventilation were in low altitude regions subject to higher temperatures compared to higher altitude areas. Consequently, the adoption of an open mound by termites might serve as an adaptation that facilitates heat dissipation in regions characterized by high temperatures. This observation suggests that structures primarily designed for gas exchange might also contribute to nest thermoregulation when favourable thermal conditions are available (Perez & Aron, 2020).

Hence when looking at temperature in termite mounds, it is crucial to consider factors like mound size and ventilation. More comparative studies on temperatures in inhabited and uninhabited mounds could provide valuable insights.

### (3) Effect of soil temperature on nest temperature

Temperature regulation by termite mounds is still a matter of debate (Woon *et al.*, 2022). Some researchers argue such a regulation does not occur (Turner & Soar, 2008). Studies on *M. michaelseni*, a species with an underground nest, revealed that the nest temperature is strongly correlated with the soil temperature at 1 meter depth. Hence any difference in temperature between the soil and the nest would lead the heat to flow in a direction to facilitate heat gain. Thus, according to these authors, the high heat capacity of the soil contributes to daily temperature damping and insulation, preventing abrupt temperature fluctuations within the nest. However, the soil may not offer sufficient temperature conservation throughout the year, resulting in more variations in the nest temperature, typically around 17°C. They also reported that the mound does not actively contribute to temperature regulation but instead disrupts the temperature within the nest as it is more affected by the ambient temperature.

On the contrary, other authors claim that that the mound alone is capable of maintaining a constant nest temperature (Korb & Linsenmair, 2000). In the *T. trinervoides* species, whose nest was at the central part of the mound just at the soil interface referred to as core (**Figure 3**), the primary source of thermal insulation was attributed to the mound itself due to the absence of an underground nest. The epigeal mound maintained a constant nest temperature and protected it from solar irradiation and ambient temperature fluctuations. The mound provided temperature buffering against external temperature in the same capacity as though the nest was embedded 20 mm below the soil (Field & Duncan, 2013). The same observation was noticed in the *Coptotermes lateus* species, also having an epigeal nest, where the core temperature at the nest did not track those of the soil (French & Ahmed, 2010).

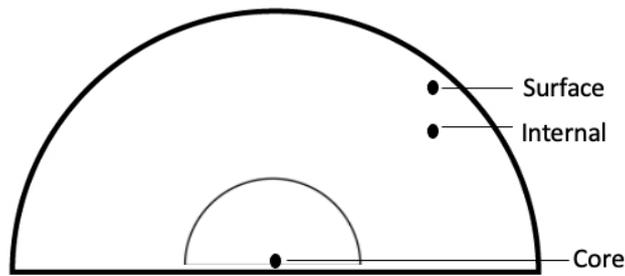

*Figure 3. A sketch showing different parts of a mound (adapted from Field & Duncan, 2013).*

In a study where the mound was damaged because of measuring instruments, the nest temperature became unstable and began to decline until the termites repaired the mound (Field & Duncan, 2013). This observation highlights the temperature regulation in the mound in an epigeal nest, where the effect of the soil is negligible. The situation may be different for subterranean nests, which requires further investigations.

It is clear that the mound plays a major part in regulating temperature in the nest of epigeal mound. It is important to carry out further studies in order to find out the influence of soil buffering capacity on nest temperature especially in subterraneous nests.

**(4) Effect of ambient temperature on nest temperature and structure.**

The examination of temperature profiles in termite nests has revealed that irrespective of fluctuations in ambient air temperature, temperature remains relatively constant suggesting that it is effectively regulated. It is important to highlight that the baseline for the nest temperature is determined by ambient temperature (Korb & Linsenmair, 2000; Korb, 2003). This phenomenon was demonstrated in studies in *M. bellicoscus* where higher ambient temperatures led to higher nest temperatures, and vice versa. This relationship was further confirmed through the manipulation of the local environmental temperature around the nest, by providing shade to the termite mounds with grasses, which reduced the surrounding temperature and in turn caused a decrease in the nest temperature.

In addition to its impact on nest temperature, ambient temperature also affects the architecture of the mound (Perez & Aron, 2020). Mound architecture is an adaptation to long-term changes of the environmental conditions, particularly temperature (Aiki *et al*., 2019). High ambient temperatures lead to high nest temperatures, causing the termites to modify the structure in order to increase the dissipation of heat. In contrast, in cooler regions, the architecture of the mound is altered to reduce heat loss (Korb, 2003). **Figure 4** shows some examples of the diversity of mound architectures built by termites of different species as well as the same species.

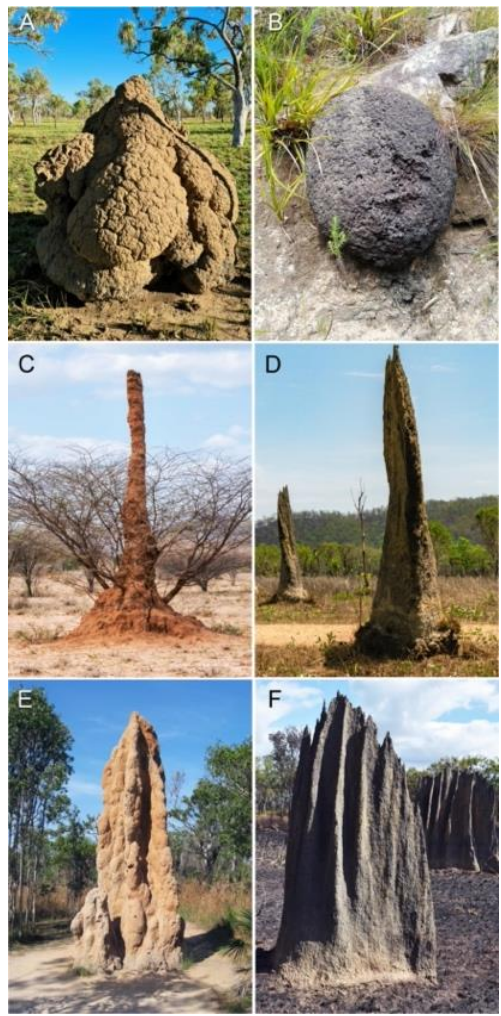

*Figure 4. Diversity in termite mound architecture (A) Nasutitermes triodiae in Western Australia, (B) Amitermes hastatus South Africa (photograph by Nicole Loebenberg), (C) Macrotermes jeanneli in South Ethiopia (photograph by Rod Waddington), (D) Amitermes meridionalis in Northern territory, Australia (photograph by Peter Szoke), © Nasutitermes triodiae in Northern Territory, Australia (photograph by John Brew) and (F) Amitermes laurensis in Queensland, Australia (photograph by Russell Cumming).*

Figure 4A and 4E show that the mounds built by *Nasutitermes triodiae* at different stages of development have very different architecture. This implies that within the same termite species, the surface area, internal structure as well as the external structure can be very different. These differences in architecture have a direct impact on temperature regulation and the ventilation process. The mounds built by the same termite species can have different ventilation mechanisms even in the same climate. Similar architectural diversity is observed in mounds belonging to the *Amitermes* genera (Figure 4B, 4D and 4F). These differences in architecture in the same species, adds an extra layer of complexity while studying termite mounds. The fact that the same termite species can build mounds whose architectures can be very different suggest that the building behaviour of termites can be modulated by the materials properties used to build the nest and environmental factors like temperature, humidity, or $CO_2$ levels (Claggett *et al.*, 2018; Perna & Theraulaz, 2019; Yang *et al.*, 2022).

In regions characterized by high solar irradiation, the mounds exhibit a propensity towards conical shapes, with a noticeable inclination towards the direction of the sun. Conversely, in areas characterized by lower ambient temperatures, dome-like structures are more prevalent (Fagundes, Ordonez, & Yaghoobian, 2021). This alteration of architecture is also observed in *M. bellicosus* that build cathedral shaped nests in the savanna and dome shaped nests in the gallery forest. The savannah region, characterized by scattered trees and open grassland, experiences high temperatures due to abundant sunlight, while the gallery forest, composed of dense trees forming a canopy, experiences lower temperatures (Woon *et al.*, 2022). The cathedral-like shapes have thin walls, and high surface complexity (the ratio of the surface area of the mound to that of an ideal cone with same basal height and width (Korb & Linsenmair, 1998)), with nest temperatures around 30 °C, which is optimal for fungus cultivation. In contrast, the domed mounds have thick walls and low surface complexity, with nest temperatures close to 28 °C, reflecting the need to conserve minimal heat from the sun.

The impact of the wall thickness on thermal properties has not been studied; however, we suggest that the thermal properties of the mounds are significantly influenced by wall thickness. This is evident in the savannah nest, which has thinner walls and less insulation, while the gallery forest nest exhibits the opposite trend.

Limited research has been conducted on the influence of mound architecture on its thermal properties although our current understanding points to the fact that mound architecture largely influences mound temperature. Conducting additional studies in this area is crucial as it will provide a comprehensive understanding of temperature regulation mechanisms within the mound.

**III.     Ventilation in termite mounds**

Termite nests require effective ventilation to exchange $CO_2$ produced both by the termites and the fungi they house. Lüscher (1961) estimated that the oxygen requirement for a mound containing a colony of approximately 2,000,000 termites weighing 20 kg to be approximately 240 L/day, which is 1,200 L of air per day. The gas output of large mounds like the *Macrotermes jeanneli* is between 100,0–0 - 400,0000 L air per day including 800–1,500 L $CO_2$ per day (Darlington *et al.,* 1997).

The presence of approximately one million workers in a colony of *Macrotermes* requires an oxygen supply comparable to that required by large mammals, such as goats or cows. Typically, the oxygen concentration inside a nest is 2% lower than the surrounding air, while carbon dioxide levels are higher up to 6% (Turner, 2005; Ocko *et al.,* 2017). These values indicate the importance of ventilation in termite nests.

In termite mounds ventilation happens in two stages (King *et al.,* 2015; Abaeian, Madani, & Bahramian*,* 2017). The first stage is transport of gases from the underground part of the nest, where metabolic activity takes place (*i.e.*, where the fungus and the brood are located), to the mound surface. In the mounds that do not have an underground part, this first stage consists of transporting the gases from the core of the nest to below the outer wall of the mound. The second step is the transport of gas across the outer wall of the mound to the outer environment. The outer wall of the mound been reported to be porous, which can contribute to ventilation (Schmidt, Jacklyn, & Korb, 2014; Abou-Houly Eid H., unpublished data; Singh *et al.,* 2019).

The presence of an interconnected network of open channels is also important to facilitate the distribution of air within the nest. One can distinguish three types of network structures within the mound: (1) the tunnels around the underground nest (when it is present) that have a large diameter, (2) the middle network of tunnels connecting the underground section to the upper section of the mound (chimney) and then (3) the surface conduits and tunnels. These tunnels can have a narrow diameter which helps in capillary action or a large diameter that actually help in air flow (Turner & Soar, 2008).

Diffusion can help stabilize the gas gradient across the mound, but it may not be sufficient to transport the gas across the nest surface. The gas requires four days to diffuse across 2 meters which is equivalent to 0.02 m in 1 hr (King *et al.,* 2015). Most large nests cannot rely on diffusion but on bulk flow in the nest because the ventilation that results from diffusion would not be effective. However, in magnetic termite mounds such as *Amitermes meridonalis* diffusion is the main source of ventilation (Schmidt *et al.,* 2014). This is a consequence of the large surface area of the mound especially in food storage areas (see **Figure 4**D). Hence, diffusion may play an important role in ventilation depending on the mound type. It is however important to mention that the colonies of *A. meridonalis* are smaller compared to fungus farming termites, thus the size of mound should be taken in consideration when examining gas transport in the mound. On the other hand, bulk flow is mainly driven by internal temperature changes (i.e., the thermal buoyancy within the nest) or external wind (Korb & Linsenmair, 1999a; Turner, 2001).

There have been several theories on how gas is exchanged in termite mounds in both closed and open mounds. Most of these theories have not been conclusively proven due to the difficulty in measuring and understanding the circulation of air flow in the mounds (King *et al.,* 2015). We discuss some of these theories in the following sections.

**(1) Thermosiphon theory and unidirectional ventilation**

One model that has been put forward to explain ventilation in a termite mound relies on the thermosiphon theory. In this case ventilation is created by the upward movement of warm air from the nest and the downward flow of cool air through surface ridge air channels, driven by differences in density creating a loop (Lüscher, 1961). **Figure 5**A shows a representation of the thermosiphon process in the mound. In a closed mound housing a large colony of about two million termites, the heat resulting from the metabolism of the termites and fungus (if present) may reach up to 100 watts. This heat produced flows up from the nest as hot air moves through the central shaft to the top of the mound driven by convection (buoyancy). Then, this air moves towards the ridges of the surfaces that have a lower temperature. Fresh air from the outside of the mound is then absorbed via channels in the ridges as cooler air descends towards the nest leading to a cyclic ventilation process.

In open chimney mounds, a better ventilation may take place as a result of the venturi effect or a stack effect. Stack effect describes ventilation in tall mounds where air enters via the vents at the bottom and escapes via vents from the top due to lower pressure at the top of the mound than the bottom. This unidirectional ventilation has been reported to occur in *M. subhyalinus* and *M. jeanneli* (Darlington J., 1987; Darlington J., 1989).

The thermosiphon theory effectively explains ventilation processes in larger closed mounds; however, the question arises as to how ventilation is initiated in smaller mounds with limited heat production.

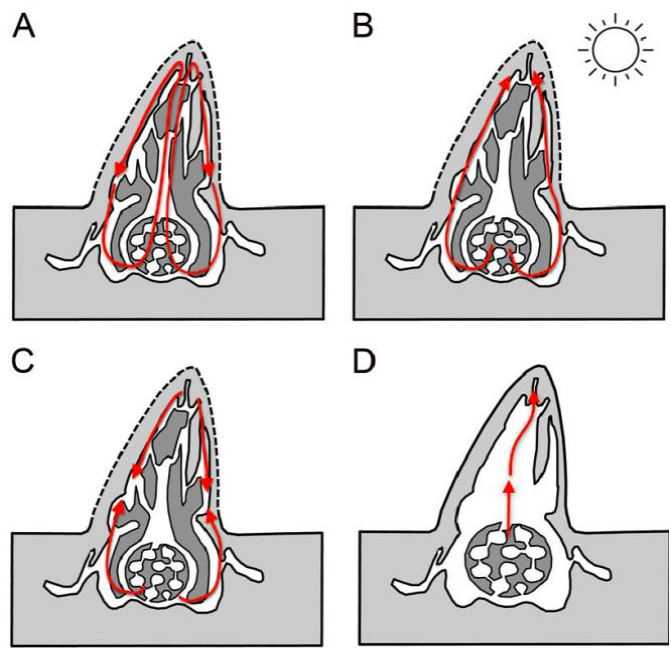

*Figure 5*. *Ventilation of Macrotermes bellicosus mounds. The red arrows indicate the direction of air flow. (A) Thermosiphon ventilation mechanism as proposed by M. Lüscher. (B) Cathedral shaped termite mound in the savanna of northern Ivory Coast during the day (external driven ventilation). (C) Cathedral shaped termite mound in the savanna of northern Ivory Coast during the night (internal driven ventilation). (D) Dome-shaped termite mound in the forest of northern Ivory Coast (internal driven ventilation). For more information see text (modified from Korb, 2011).*

**(2) Internal and external driven ventilation**

Another model that accounts for the dynamics of air flows in termite nests is the internally and externally driven ventilation. This model was proposed by Korb and Linsenmair (1999) after testing the thermosiphon model postulated by Luscher (1961).

In the external driven ventilation model, the ventilation occurs during the day and is mainly driven by ambient temperature. According to this model, solar heating increases the air temperature in the channels inside the ridges through the mound surface, allowing the warm air to rise and respiratory gases to be exchanged through the thin walls in this process. In this case, $CO_2$ concentration is low due to increased diffusion rate as result of elevated temperature, increased area for gas exchange and circulation of air current.

At night and at all times in dome-shaped mounds, the internally driven ventilation occurs. During these times, the nest has a higher $CO_2$ concentration that is consistent in all directions. As ambient temperature is lower than the air channel temperature, which is in turn lower than the nest temperature, air rises from the nest, facilitating gas exchange through the central shaft and external surfaces. The internal ventilation mechanism occurs in cooler environments and has a lower efficiency compared to the external ventilation mechanism as the $CO_2$ concentration was found to be consistently higher during the night. The internal and external ventilation mechanisms are depicted in **Figure 5**B-5D.

The internal driven ventilation is similar to the thermosiphon model proposed by Lüscher (1961), where air flow is driven by the higher temperature of the nest and the internal parts of the mound. However, unlike the thermosiphon model, air does not return to the nest from the external ridges in a perfectly circular or loop system. The airflow patterns in both the internal and external driven models, which deviate from perfect circular motion, could arise due to either a temperature gradient hindering the flow or a lack of adequate peripheral air channels. In the case of the later, this suggests that the type of ventilation in a termite mound is influenced by the structural components present in the mound. For instance, in a mound that has thick walls and if the external temperatures are high, it provides room to support externally driven ventilation. However, the absence of thin walls hinders the transfer of heat into the air channels within the mound, limiting this process.

The driving force behind externally and internally driven ventilation, which is based on diurnal temperature changes, closely aligns with the solar-driven ventilation which was later suggested by King *et al.* (2015) and Ocko *et al.* (2017) after conducting direct measurement of air flow in the mounds. This solar powered ventilation may be the most accurate mechanism proposed to describe ventilation in termite mounds and is likely to be generic for all mound building species, as daily oscillations and sunlight are ubiquitous in the environments where termites are found. However, variations may occur due to local environmental factors as well as the mound architecture itself. For instance, in mounds with thicker walls, there may be a higher $CO_2$ concentration in the mound that may result from poor $CO_2$ exchange across the walls. This phenomenon highlights the importance of mound architecture and surface properties for ventilation. Therefore, it is crucial to examine the physical properties of the walls to determine their impact on ventilation.

### (3) Tidal effect

A different model put forward to explain ventilation in termite nests rely on the tidal effect. According to this model, wind is a crucial factor in controlling air movement in the mound, as it influences ventilation through its speed and direction (Turner & Soar, 2008). Lüscher's concept of thermosiphon was found to be insufficient in explaining the ventilation mechanisms present in complex mound structures such as those of *M. michaelseni* (Turner, 2001).

In this model, the observed movement of air in the mound was too complex and strongly driven by wind in contrast to the circulatory air movement suggested by Lüscher (1961). In addition, the air movement suggests an interaction between metabolism-induced buoyant pressure in the chimney and the wind-induced pressure in the surface conduits, with the lateral connectives serving as a damping mechanism to create a balance between the two forces.

The tidal effect explains how wind affects ventilation in termite mounds, but it does not explain how ventilation occurs in the absence of wind. Does this suggest an absence of ventilation during these periods? It is important to highlight that the effect of wind on the mound was later discovered to be transient as wind velocity around the nest was small and only significant on very windy days or windy moments (Ocko *et al.*, 2017).

In summary, the ventilation process is crucial, especially in closed mounds with no apparent holes. Open mounds, on the other hand, have an easier ventilation process due to small millimetre holes on the mound. For instance, the presence of egress compress (a network of reticulated channels that form on the spire of the mound) during certain seasons on the mound of *M. michaelseni* nests has been observed to favour in gas exchange between the interior and exterior of the mound especially during rainy season (Abou-Houly Eid H. unpublished data; Andréen & Soar, 2023).

It would be interesting to study in more detail the ventilation mechanisms in smaller mounds and those without a developed central shaft or complex surfaces. It remains unclear if solar power alone drives ventilation in such environments and how the absence of circular interior surfaces may impact air circulation. Additionally, most measurements in larger mounds have been taken at the peripheral surface and nest, leaving the boundary between upward and downward air flow patterns in the interior of the nest undefined.

**IV. Combined effect of ventilation and thermoregulation in mounds**

As discussed in the previous sections, termites modify their mounds to adapt to environmental changes, and these modifications directly impact both ventilation and thermoregulation. Therefore, it is important to examine simultaneously the thermoregulation and ventilation processes within a termite mound and how they interact and influence each other. For instance, in cold environments, termites construct dome-shaped mounds to minimize heat loss, resulting in thicker walls that hinder gas exchange (Perez & Aron, 2020; Fagundes *et al.,* 2021). Prolonged obstruction of air movement can lead to air saturation and even condensation within the pores in the mound walls (Lüscher, 1961). The increased $CO_2$ levels in the mound further causes a reduction in termite metabolic activity, inducing dormancy. The consequence of the reduced metabolic heat production contributes to a further decrease in nest temperature, contrary to the termites' initial intention of conserving heat (Korb, 2003).

As a matter of fact, the design of termite mounds influences three concurrent processes: humidity, ventilation, and temperature regulation. However, it is difficult to determine whether these properties respond directly when they reach excessively high or too low levels, rather than just being affected by coincidence (French & Ahmed, 2010).

Ventilation is changed by factors such as surface area available for gas exchange, wall thickness, and airflow patterns within the mound. The impact of reduced ventilation due to thermoregulation is more pronounced in larger termite colonies because they have higher oxygen requirements. In contrast, smaller colonies can better tolerate reduced ventilation as they need less oxygen due to their smaller population and less competition for oxygen. The interplay between ventilation and thermoregulation ultimately determines the size and distribution of a particular termite species' mounds. It is noteworthy that only a few termite species, primarily found in Africa, have evolved mechanisms to maintain adequate ventilation through thick walls (Lüscher, 1961).

Nest temperature exhibits an exponential increase with ambient temperature, as the latter directly influences both nest temperature and metabolic heat production. Whereas, the $CO_2$ concentration in the nest shows a linear increase with ambient temperature, as higher temperatures stimulated nest metabolism, leading to increased $CO_2$ production.

In order to adapt to changes in ambient temperature, the surface of the mound needs to be exponentially reduced in low ambient temperatures to minimize heat loss and maintain nest temperature. Conversely, as ambient temperature increases, the surface mound must increase linearly to reduce $CO_2$ concentration (Korb & Linsenmair, 1999a). This adjustment of surface of the mound aims to keep the nest within the optimal and viable ranges for colony survival, as illustrated in the model presented in **Figure 6**.

Every mound is primarily predisposed to carry out either thermoregulation or ventilation as these two processes maybe conflicting. Consequently, the primary function of the mound depends on who inhabits the nest (termites and other organisms), if the mound is above or below the ground and the environmental conditions (Korb, 2011). By studying the interrelationships between thermoregulation and ventilation in termite nests, we gain valuable insights into the complex dynamics of these processes and their impact on termite colonies.

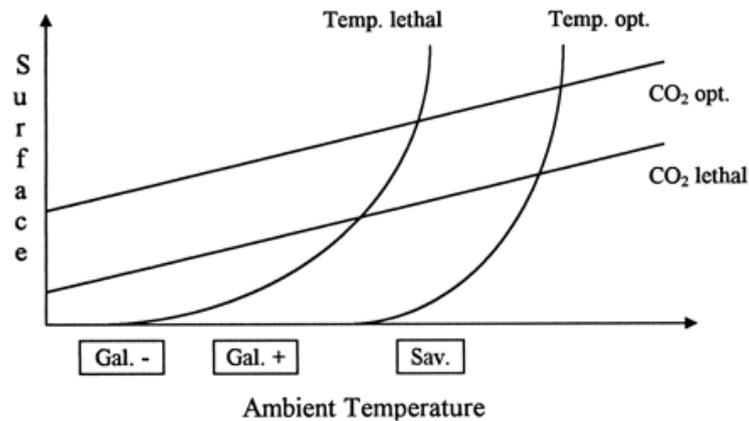

*Figure 6. Interaction between surface area, ambient temperature, and $CO_2$ concentration (adapted from (Korb & Linsenmair, 1999a).*

**V. Humidity**

In addition to specific thermal requirements within their nests, termites, and the fungi they cultivate also require high humidity conditions. Such conditions are necessary because termites are prone to desiccation due to factors such as the scraping off their thick chitin outer layer (cuticle) during tunnelling or improper development of their outer surfaces. This thick chitin layer would have also offered protection against temperature fluctuations if present (Claggett *et al.,* 2018).

To maintain the required moisture levels, termites employ strategies such as cultivating fungi within the nest or digging outside the nest to access water from underground sources, which is then brought back into the nest (Lüscher, 1961). The regulation of water within the termite mound is a process that involves a balance between water inflow and outflow. In the case of *Macrotermes* mounds, water flux circulates along four distinct pathways: the evaporation from the mound's surface, the passive movement of water from neighboring soil into the nest, the active transportation of water to the nest through termite-carried salivary glue, and the active transportation of water from the soil located beneath (Turner, 2006).

Termites detect humidity with their antennae and have the ability to distinguish between areas of high and low humidity levels (Bardunias *et al.,* 2020). In studies conducted on five species in Ivory Coast, long-term measurements of nest humidity consistently revealed values ranging from 98% to 99% (close to saturation point), with the lowest recorded value of 96.2%. However, lower humidity levels around 70% were found in the nest of *M. michaelseni* in Namibia. These values are however consistently higher than the external humidity. Moreover, humidity is comparatively lower in the chimney, suggesting that gas exchange occurs in that area.

Termites may also adjust mound structure based on humidity changes (Turner, 2001). For instance, experiments in low humidity conditions, *Macrotermes michaelseni* showed reduced soil transport and built smaller structures, also altering their microstructure (Carey *et al*., 2019). The reverse has been observed in high humidity conditions, where termites expand the construction site facilitating larger mound development (Bardunias *et al.,* 2020). This adaptation to humidity conditions is also essential for creating optimal environment for farming fungi.

The mechanisms underlying ventilation in termite mound particularly in the presence of high humidity, raise several questions. Considering that the diffusion of gases through water is approximately 1000 times slower than through air, one may ask how these nests achieve adequate ventilation despite the low diffusion rates. When there is a slow diffusion rate, it takes two days for the gas to diffuse through a distance of 4 meters. If one considers the further reduction in diffusion caused by high humidity, this suggests that diffusion may not be the dominant mechanism in most nests, compared to advection propelled by temperature and pressure gradients. Ventilation may be driven by advection, controlled by the diurnal cycle as proposed by King et al. (2015). Additionally, high wind velocity can propel advection in the mound coupled with the presence of large micro-scale pores in the mound walls (Singh *et al.,* 2019).

To gain a comprehensive understanding of the combined effects of humidity and thermoregulation on ventilation, it is crucial to conduct experiments that replicate these conditions in both fungus and non-fungus farming nests.

**VI. Investigating the architecture of termite mounds with X-ray tomography**

Empirical evidence has demonstrated the ability of termite nests to regulate temperature, humidity and to facilitate efficient gas exchange. Nevertheless, the precise roles of structural elements such as the walls of mounds, their micro-structures or the air channels on gas flows and the thermoregulation process are still not well defined. Furthermore, the manner in which flow properties differ across scales, ranging from larger to micro dimensions, remains poorly understood.

A first step to gain a deeper understanding of the functional properties of termite mounds architectures can be achieved by imaging their internal structures. Historical methods used to visualize the interior of mounds included cement infiltration, mound sectioning and paint tracing, cross-sectioning image analysis, planar sectioning, slicing and scanning (Darlington, 1985; Turner, 2000; Nauer *et al.,* 2018; Abou-Houly Eid H. unpublished data). Each of these techniques had its own strengths and weaknesses. However, many of these methods failed to provide a clear and comprehensive representation of mound's internal features. Additionally, these approaches are destructive, rendering the mounds unusable for other purposes after acquiring images. Electric resistivity tomography (ERT) has also been used to study the internal structure of the mound. Although non-invasive, this method only provides surface electric resistivity values and does not offer a clear image of the mound's internal structure, posing a limitation (Van Thuyne *et al.,* 2023).

X-ray tomography provides a means of visualising the internal structure of materials non-destructively. This technique has the capability to generate reliable three-dimensional representations of materials, even those with opaque properties (Perna et al., 2008; Wildenschild & Sheppard, 2013). The principle is based on the attenuation of X-rays as they pass through a material, which follows the Lambert-Beer Law (Tapias Hernández & Moreno, 2020). The configuration of X-ray computed tomography (CT) scanners typically includes a source and a detector, which either rotate around the sample, as in medical CT, or are stationary and the sample rotates on a rotation stage, as in laboratory micro-CT. The sample is positioned between the source and the detector, and the resultant radiographs are captured by the detector and then reconstructed to create 3D images (Bultreys, De Boever, & Cnudde 2016; Zhan, Yong, & Zhang, 2019).

A medical CT scanner can achieve a spatial resolution of approximately 250 µm voxel, suitable for scanning large samples as displayed in Figure 7. With advancements in technology, higher-resolution tomography techniques such as micro-CT and nano-CT have emerged, allowing for the visualization of finer details within a sample (Cnudde & Boone, 2013). The combination of different X-ray scanners can allow multi-scale characterisation of materials (Bultreys *et al.,* 2016).

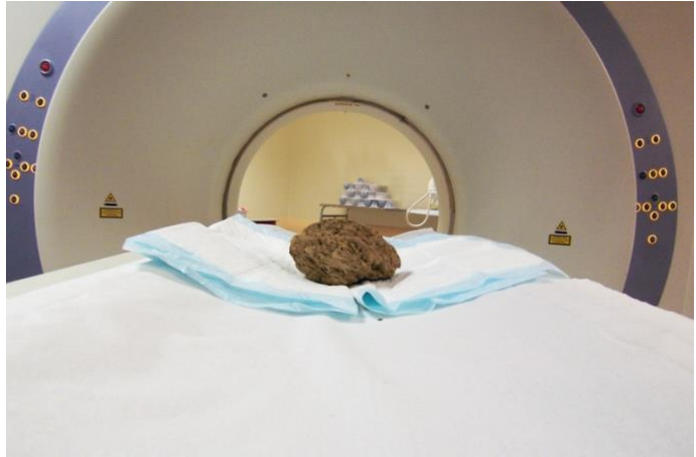
*Figure 7. Termite mound in a CT scanner.*

Although, X-ray computed tomography presents numerous benefits, it comes with a few drawbacks, such as the potential presence of noise that can interfere with the data quality. However, effective filtering techniques can mitigate this noise. Additional disadvantages encompass operator influence and imaging artefacts challenges. Notwithstanding these hurdles, X-ray tomography remains an indispensable imaging tool with diverse applications spanning various fields (Cnudde & Boone, 2013).

In the 1980s, X-ray tomography found initial applications in laboratory settings, primarily for assessing fluid saturations in soil sciences and the petroleum industry (Petrovic, Siebert, & Rieke, 1982; Vinegar *et al.*, 1987), achieving a spatial resolution of approximately 1-3 mm. The use of micro-CT for porous media imaging emerged with Exxon Research pioneering resolutions as fine as 1 µm (Flannery *et al.*, 1987).

The tomography data can be used for porous media characterisation, such as pore size distribution analysis, fracture assessment, multiscale imaging, ore evaluation, fluid flow analysis, and the monitoring of structural dynamic processes (Cnudde & Boone, 2013; Zhang *et al.*, 2019). Many of these benefits can potentially offer insights to understand the internal structure of termite mounds which can be useful to study thermoregulation and ventilation within their mounds. For instance, obtaining the channel size distribution and the permeability in termite mounds can enhance the comprehension of how flow is influenced by the spatial organization of the conduits, channels, and pores. Furthermore, the application of multi-scale imaging can facilitate an understanding of how smaller-scale processes impact larger-scale phenomena and how properties vary across scales. Moreover, X-ray tomography can be used to study fluid drainage processes in termite mounds. Additionally, direct numerical simulations can be performed on the obtained images to deduce fluid flow characteristics within the nest.

**(1) Applications of X-ray tomography to termite mounds**

X-ray tomography has been previously employed to characterize and quantify the internal structure of termite mounds. In 2001, medical CT was effectively used to visualize termite mounds, enabling a comparison between void spaces and constructed materials within the mound (Perna & Theraulaz, 2019). Additionally, Perna *et al*. (2008) employed CT to obtain a 3D network of galleries within the *Cubitermes* mound which was later used in developing a model for growth of transportation network in termite mounds by Eom *et al*. (2015). Their study revealed a compromise between the level of connectivity and defence against predators in the mound, along with ongoing rearrangement of connections throughout the mound's lifespan. Medical CT scanning has also been used to visualise the complex network of egress in *M. michaelseni*. The images have been used to study mass transfer within the mound (Andréen & Soar, 2023). Furthermore, Nauer *et al*. (2018), Van Thuyne *et al*. (2023), Zacharia *et al*. (2020b) used X-ray tomography to determine the porosity distribution in termite mounds, and Fuchs *et al*. (2004) used X-ray tomography to study mound construction behaviour in *Crypotermes secundus* hill, a species recognized as a termite pest.

**(2) Millimetre scale and microscale X-ray tomography of termite mounds**

In this section we examine termite mounds scanned at both the millimetre and microscale.

**(a) Millimetre scale X-ray tomography**

As mentioned in the previous sections, there are several X-ray scanners available in the market, each with its own advantages. When medical X-ray CT scanners are used to scan termite mounds, they are able to capture extensive volumes of the mound, and in some instances, the entire mound, as shown in **Figure 8**. **Figure 8** shows the subterranean mounds of *Apicotermes lamani* from the Republic of Congo and both epigeal and subterranean parts of mounds of the *Cubitermes sp., T. geminatus* and their corresponding virtual cross-sections. The virtual cross-section reveals distinctive circular tunnels in the *A. lamani* arranged in floors and regular patterns (Jost, 2021). The network of chambers in the *Cubitermes* mound is also apparent. Each of these mounds show unique patterns and channels, representing spaces where the termites move, as well as the solid parts of the mound. **Figure 9** shows a zoomed in image of the channels and solid in the *T. geminatus* species.

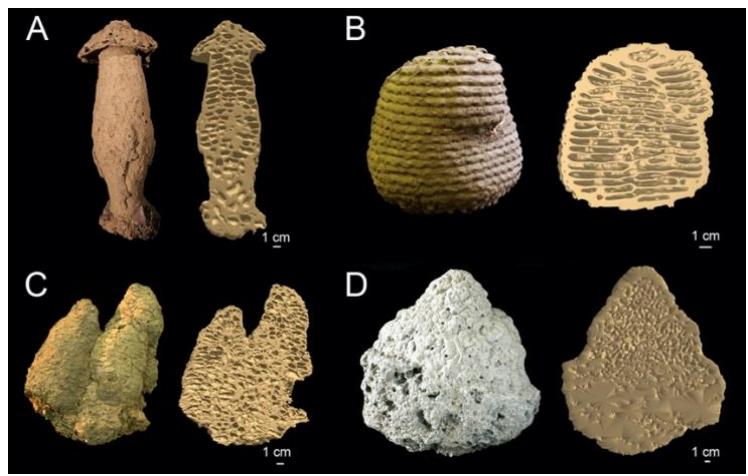

*Figure 8. Termite mounds scanned with a medica CT scanner. (A) Cubitermes sp. nest from the Central African Republic and its virtual cross-section. (B) Subterranean nest of Apicotermes lamani from the republic of the Congo and its virtual cross-section. (C) Procubitermes sjoestedti from Ivory Coast and its virtual cross-section (D) Trinervitermes geminatus from Guinea and its virtual cross-section (MESOMORPH Project, CRCA, CNRS, Toulouse).*

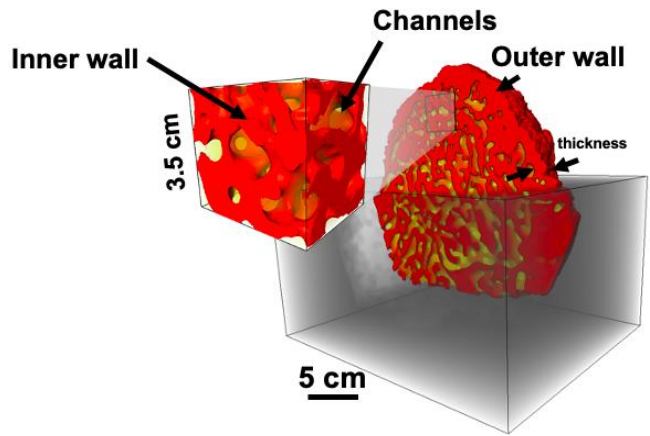

*Figure 9.* *CT scan of the termite mound, T. geminatus, showing channels and walls* (adapted from Singh *et al*., 2019).

For exceptionally large mounds that cannot be scanned in their entirety, it is possible to cut them into small pieces and scan them individually. Subsequently, these individual scans can be merged together to generate a complete image of the entire mound. An example of this method is demonstrated in the scanned mound presented in Figure 10 obtained from our fieldwork conducted in Nigeria.

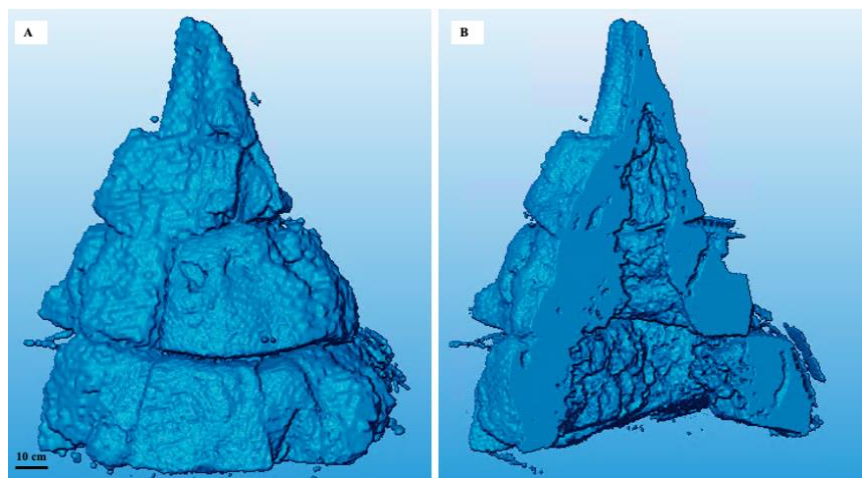

*Figure 10.* *Assembled whole mound A) Macrotermes* sp *mound from Port Harcourt, Nigeria and (B) its vertical cross section.*

CT tomography provides a broad view of the termite mound, allowing us to capture distinctive features of each mound's structure and architecture. This imaging is particularly valuable as it aids in understanding the correlation between large-scale features, such as inner and outer wall thickness, overall structure, and the mound's functionality. Additionally, this imaging offers insights into the arrangement of chambers and channels without destroying the mound.

**(b) Microscale X-ray tomography:**

Microscale tomography of termite mounds provides enhanced resolution, achieving a voxel size of 1 µm (or even further down), depending on the machine's capability (Singh *et al.,* 2019). This high resolution facilitates the detailed visualization of small-scale features, including micropores and metallic elements in the walls, and even cracks on the walls, as shown in **Figure 11**.

While microscale tomography has proven popular and useful in various fields (Blunt *et al.,* 2013, Alhosani *et al.,* 2021; Wang, Spurin, & Bultreys, 2023; Goodarzi *et al.,* 2024), its application to termite mounds has been reported in only a few papers including (Singh *et al.,* 2019; Oberst *et al.,* 2021; Oberst and Martin, 2023). The drawback of microscale tomography is its limited field of view, allowing for the imaging of only small portions of the mound. Typically, to obtain a comprehensive view of the entire mound structure, a combination of microscale and macroscale tomography is required, providing a comprehensive view with finer details.

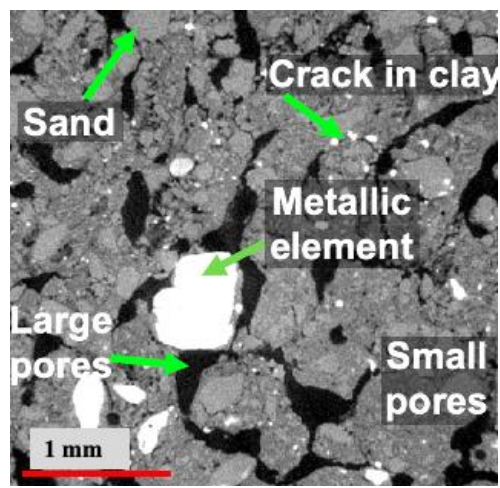

***Figure 11.*** *X-ray micro-CT image of T. geminatus* (adapted from Singh *et al*., 2019).

**(3) Image processing of CT and micro-CT mound images**

To get the most out of tomographic analyses, images must be processed by filtering and segmenting. To enhance the image quality and reduce noise, a filter is generally applied to grey-scale images. Non-local means filter has become popular in this regard and provides good image quality (Buades, Coll & Morel, 2005; Buades, Coll & Morel, 2008; Li *et al*., 2014; Zhang *et al*. 2017; Youldash *et al*., 2021). **Figure 12**A and 12B provide an example of an image before and after this filtering process.

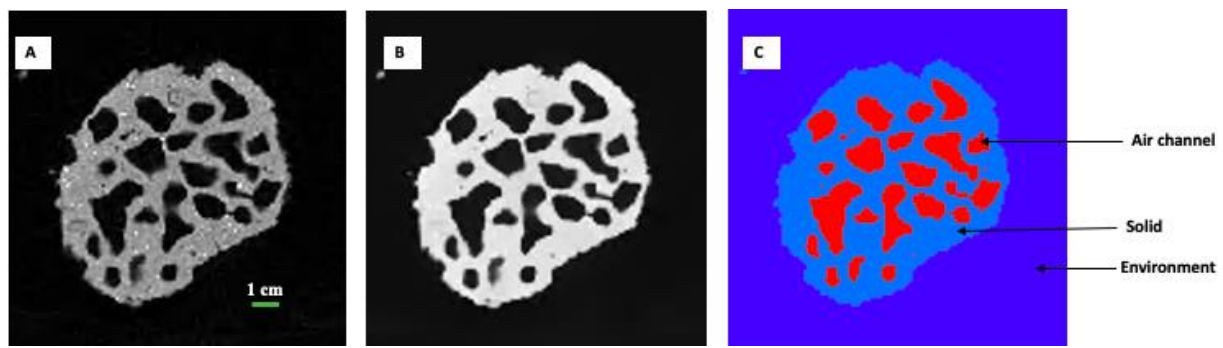

*Figure 12. Different stages of image processing. (A) unfiltered image (B) filtered image (C) segmented image.*

The filtered images can be segmented into distinct phases: (1) the solid areas, which constitute the sections constructed from dense mound building materials, and (2) the channels, which represent air spaces or pores that termite use to move within the mound. An example of a segmented image is shown in **Figure 12**C. The microscale images that have more details, of e.g., mound walls, can be segmented into various phases, e.g., pore space, grains, and metallic elements. The pore space can further be divided into individual pores. Segmentation can be accomplished using a machine learning based algorithm, called Weka, within Fiji software (open-source software). Alternatively, other open-source (such as Dragonfly) or commercial software can also be used for this purpose. Details about these software is shown in Table 1. Once the segmentation is complete, these images can be used for further analysis to assess parameters like porosity, wall thickness, connectivity and perform flow simulations.

*Table 1. Useful open-source software for image processing and simulation*

| S/N | Name of software | Use | Source |
|---|---|---|---|
| 1 | Fiji | Machine learning tool for segmenting 3D images | https://imagej.net/software/fiji/downloads |
| 2 | Dragonfly | Segmentation of 3D images | https://www.theobjects.com/dragonfly/index.html |
| 3 | OpenFoam | Open-source tool for computational fluid dynamics (CFD) | https://www.openfoam.com/current-release |
| 4 | GeoChemFoam | Open-source tool based on OpenFoam CFD toolbox | https://github.com/GeoChemFoam |

**VII. Numerical simulations on digital models and 3D structure of termite mounds**

A few studies have used flow simulation techniques to investigate the ventilation and thermoregulation properties of termite mounds. Most studies on termite mound have been carried out *in situ* rather than developing models that effectively capture the underlying physical processes taking place within the mound architecture. Integrating numerical simulation alongside with fieldwork and X-ray tomography offers a valuable approach for analysing the intricate processes occurring within the mound in a controlled and computationally efficient manner. This integrated approach also provides a cost-effective approach for modelling a variety of fluid flows in different mound structures (Abou-Houly Eid H., unpublished data). We examine below some of the existing research using numerical simulations on termite mounds and how they can be incorporated in tomographic images of mound.

**(1) Numerical simulation in 3D models of termite mounds**

In this section, we discuss previous studies that focused on numerical simulations conducted on digital models that mimic termite mound structure.

**(a) Numerical simulation to investigate temperature distribution in the mound**

Several studies have demonstrated that the temporal and spatial variations in temperature within the termite mound can be effectively modelled using the heat equation (Gouttefarde *et al.,* 2017; Abou-Houly Eid H., unpublished data; Jost, 2021). The heat equation for a 1D system including an additional term (Q) representing either a heat source or sink is shown in equation 1.

$$\frac{\partial T}{\partial t} = D \frac{\partial^2 T}{\partial z^2} + Q \qquad (1)$$

where T represents the temperature in Kelvin, t represents time, D denotes the diffusivity coefficient in square meters per second (m²/s), and z represents the depth.

These studies incorporated *in situ* temperature measurements collected from distinct regions within the mound, the surrounding soil, and a numerical scheme to model heat flow. The range of the values of heat diffusivity coefficient obtained from some mounds was between $0.27 \times 10^{-6}$ to $0.80 \times 10^{-6}$ m$^2$/s (Gouttefarde *et al.,* 2017). These values closely resemble those observed in soils with comparable compositions. Additionally, the Q term consistently exhibits positive values, suggesting the presence of a heat source term originating from either the termites or the mound structure itself.

**(b) Numerical models to investigate the shape of termite mounds**

Numerical simulation techniques have been used to investigate the factors contributing to the diversity of shapes observed in termite mounds. One notable study is by Fagundes *et al.* (2021) which has introduced a free shape model, capturing the essential characteristics of complex mounds such as those found in *M. michaelseni* species. The model allowed the systematic morphing of a mound into different shapes under the influence of external forces until an optimized shape was reached, considering the requirements for thermoregulation and gas exchange. The optimization of thermoregulation was achieved by minimizing the difference between the underground nest temperature ($T_{nest}$) and soil temperature ($T_{soil}$) represented by a dimensionless temperature in the simulation, while gas exchange optimization involved maintaining a low dimensionless scalar concentration parameter which represents the gas concentration within the nest. The impact of solar irradiation, zenith angle and wind speed on termite mound morphology was also studied.

Their results demonstrate that despite the freedom of the models used to generate any mechanically stable structures, the resulting shapes bear resemblance to those observed in natural environments. In regions characterized by high solar irradiation, the mounds exhibited a propensity towards conical shapes, with a noticeable inclination towards the direction of the sun. Conversely, in areas characterized by lower ambient temperatures, dome-like structures were more prevalent. This observation is similar to the one reported for *M. bellicoscus* (Korb, 2003). Furthermore, in regions experiencing stronger winds, the mounds tended to be shorter in height. These observations suggest a correlation between mound shape and the specific environmental conditions within which they are located. Overall, the shape of the termite mound was found to be closely related to its regulatory function and the prevailing local environmental factors like solar irradiation, ambient temperature, and wind patterns.

Numerical simulation has also been used to investigate the response of termite mound morphology to internal odours, particularly those associated with pheromones released by the termites (Ocko *et al.,* 2019). Figure 13 shows how two different mound geometry evolves during morphogenesis. The results align with the findings that larger termite colonies necessitate a greater quantity of odour at a higher production rate than smaller colonies, resulting in the formation of mounds that have larger radii.

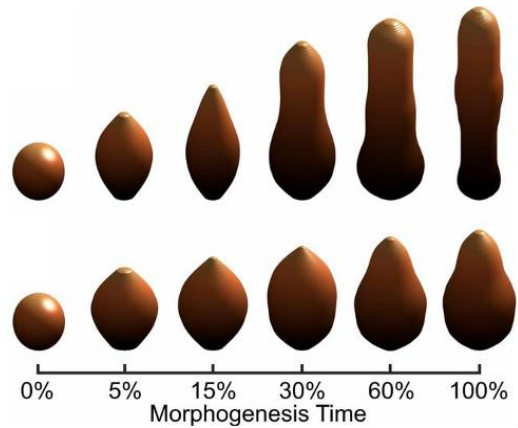

*Figure 13. Mound geometry progression during morphogenesis (Ocko et al., 2019).*

These numerical simulations in termite mound models have provided a promising starting point for modelling the processes occurring within termite mounds and have validated the use of the heat equation for modelling temperature profiles within the mounds. However, these models have primarily focused on heat and fluid flow within the mound, with limited consideration given to the internal structure and small-scale properties. With access to sufficiently detailed information about the mound's geometry and internal structure especially through X-ray tomography, a more thorough understanding of the flow patterns within the mound and their interplay with external temperature, humidity and wind conditions could be established through numerical investigations (Abou-Houly Eid H., unpublished data).

**(2) Numerical simulations on 3D tomographic reconstructions of termite mounds**

After acquiring the geometric and structural intricacies in termite mounds through X-ray tomography, these tomographic images can be used as input data into computational fluid dynamics (CFD) software for flow simulations. These simulations help us to get information about velocity fields, pressure fields, thermal conductivity and $CO_2$ diffusivity which gives us a better understanding of flow properties within the mound.

By solving the mass momentum equations, the pressure and velocity fields within the mound can be determined. These equations can be solved by applying a pressure of 1 Pa at one end of the domain and zero pressure at the other.

$$\rho \frac{\partial u}{\partial t} + u.\nabla u = -\nabla p + \mu \nabla^2 u \qquad (2)$$
$$\nabla.u = 0 \qquad (3)$$

where u is velocity, $\rho$ is density, $\mu$ is dynamic viscosity, t is time and p is pressure.

The permeability which describes the ease of flow of fluid through the porous media can be described by Darcy's equation

$$U_D = \frac{K\Delta P}{v \Delta L} \qquad (4)$$

where $U_D$ is the Darcy velocity, K is permeability, v is the kinematic velocity and L is the length of the porous medium.

Molecular Diffusivity within the mound can be obtained by solving Fick's second law. Heat conductivity can be obtained by solving heat equation until steady state is achieved.

$$\frac{\partial c}{\partial t} - D_c.\nabla^2 c = 0 \qquad (5)$$
$$\nabla.(k_\beta \nabla T) = 0 \qquad (6)$$

where c is the concentration of $CO_2$ (mol/m³), $D_c$ is the diffusion coefficient (m²/s), $k_\beta$ is the thermal conductivity of each phase (W/m/K) and T is temperature (K).

To incorporate the influence of advection caused by wind flow outside the mound, the Peclet number can be employed. This parameter quantifies the ratio between the advective transport flux and the diffusive transport flux, and is defined by the following equation

$$Pe_m = \frac{U_D L}{D} \qquad (7)$$

where D is the apparent diffusivity of $CO_2$ (m²/s).

**(a) Microscale simulation**

Microscale simulation, also known as pore-scale simulation, is a prevalent approach, widely employed in various fields, including the oil and gas industry (Blunt *et al.*, 2013; Yongfei *et al.*, 2021) and soil sciences (Yang *et al.*, 2014; Hu, Liu, & Zhang, 2018; Kirill & Gerke, 2023*)*. This technique is particularly valuable for obtaining detailed insights into mineralogical heterogeneity, thermal conductivity, relative permeability, petrophysical properties, and other factors (Blunt *et al.,* 2013; Maes & Menke, 2021; Menke, H.P., Hood, K.M., Singh, K., Medero, G.M., & Maes, J., unpublished data). The equations used in simulating fluid flow and heat transfer in subsurface rocks can be applied to the analysis of microscale images of the walls of the termite mounds, providing a comprehensive understanding of the intricate details captured at this level.

An application of pore-scale flow field simulations on segmented termite mound images was carried out by Singh *et al.* (2019). These flow field simulations were conducted using an open-source computational fluid dynamics (CFD) software known as OpenFOAM (Muljadi *et al.*, 2016). Information on this tool is found in Table 1. In their study, they investigated the function of micropores of the walls of mounds built by non-fungus farming termites of *T. geminatus.* They highlighted two categories of pore sizes in the walls, large and small micro-scale pores. The large pores were found to enhance permeability by up to two times, increase $CO_2$ diffusivity by up to eight times, provide better insulation of the mound and assist in drying of the mound thus restoring ventilation in rainy days.

These findings highlight the significance of small-scale features within the mound, such as micropores, in influencing the functional properties of the mound. It is uncertain whether the micropores will have a significant impact on the functional properties in fungus-farming termites and requires further investigation.

**(b) Millimetre scale simulations**

At the millimetre scale it is possible to perform similar numerical simulations for flow and thermal properties in the mound. **Figure 14** shows typical images of velocity and pressure fields obtained from millimetre-scale simulations. These simulations can help to understand the influence of large-scale structural features like the channels and walls of the mound on its flow and thermal properties.

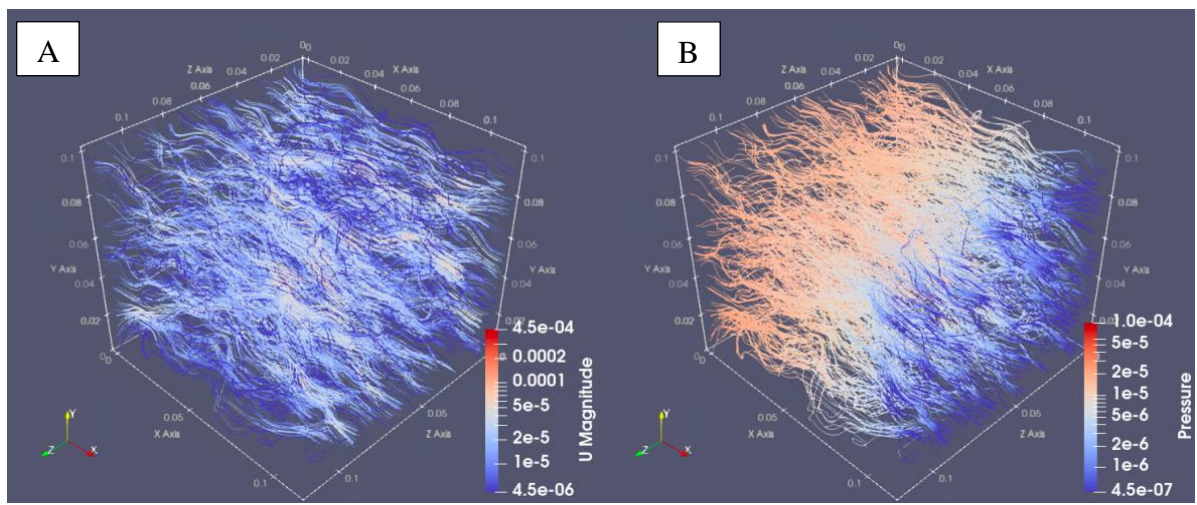

*Figure 14. Flow fields in millimetre-scale simulations performed on mound samples of T. geminatus species. (A) Velocity fields, and (B) Pressure fields.*

To get a comprehensive understanding of flow dynamics within termite mounds, mirroring realistic conditions, it is crucial to perform multiscale simulations, integrating both millimetre-scale and microscale features of the termite mound. Techniques such as Darcy-Brinkman-Stokes (DBS) can be used for this purpose. In this approach, larger channels are modelled explicitly using Stokes, while processes in microporous areas are implicitly represented as averaged volumes (Maes & Menke, 2021; Menke, H.P., Hood, K.M., Singh, K., Medero, G.M., & Maes, J., unpublished data). This method can help to understand the influence of micropores of mound walls on mound properties. In ongoing unpublished research, our results indicate that micropores significantly reduce thermal conductivity within the mound and affect the preferred flow paths. This emphasizes the importance of conducting integrated microscale and multiscale numerical simulations, which can provide valuable insights into the mechanisms underlying gas exchange and insulation within the termite mounds. Note that these simulations have been performed on subsets of 3D images of termite mounds using GeoChemFoam. Information on this tool is found in Table 1. Similar work can be conducted on complete structure of the mound (as shown in Figure 10) which will provide us with comprehensive details of the functional properties of the mound.

**VIII. Machine learning on tomographic images of mounds**

Another less-explored approach in studying termite mounds is the use of machine learning fo pattern recognition. When directly applied to 3D tomographic images of mounds, machine learning, and particularly supervised learning, can help identify similarities within pattern structures. This type of machine learning analysis can enable us to determine if there are consistent patterns in channel or solid arrangements within mounds of the same species, environment, shape, age, size, or if the arrangements are entirely random. Additionally, machine learning can help us to identify patterns within the mound structure associated with gas exchange and temperature regulation. This identification process can help to create models that predict gas exchange and temperature regulation within the mounds. Furthermore, incorporating machine learning enhances feature recognition within termite mounds, enabling the identification of various chambers, tunnels, or materials within the mounds. This feature recognition improves our understanding of mound architecture and facilitates more accurate classification.

For mound classification purpose, a residual neural network (ResNet) model can be used. ResNet is an image classification model that uses multiple layers with different sizes. Each layer has several blocks and convolutional layers, batch normalization and residual connections (He Kaiming *et al*., 2015). While using this method, the mound tomographic images can be split into training, validation, and test data sets in the best ratio. The model can be trained on several mound images except a few images. The model can then be tested for the prediction of the identity of untrained images based on the information from the trained dataset.

Gaussian Random Field (GRF) can also be used in machine learning applications in termite mounds. Using this method one can generate images of the termite mounds and textures. Oberst and Martin (2023) used GRF to reliably mimic termite mound spinodal features in the *Coptotermes lacteus* and *Nasutitermes exitiosus* termite species. They used X-ray micro-CT scans of termite mounds to extract features and optimize the GRFs. This framework can be used to engineer termite-inspired structures and investigate novel material concepts.

Given the numerous advantages of machine learning approaches, including mound classification, mound image generation, and the prediction of thermoregulation and gas exchange under various conditions, it is important to conduct more studies in this area.

**IX. Perspectives and scope of future studies on termite mounds**

Despite the existing body of research on termite mounds, several unanswered questions and knowledge gaps persist, necessitating further investigation. These unresolved aspects encompass the following areas:

- Understanding thermoregulation and ventilation in smaller termite mounds and non-fungus farming mounds or mounds lacking chimneys. Are the underlying processes similar to those observed in larger mounds?

- Examining the influence of structural properties of mounds, such as inner wall thickness, outer wall thickness, air channels, porosity, surface-to-volume ratio, and surface complexity, on gas exchange, drainage, and thermoregulation within the nests.
- Studying thermoregulatory and flow properties across various scales, ranging from the nanoscale, microscale to millimetre (large) scale. Alterations in flow properties across these scales are not currently well understood.
- Analysing the combined impact of mound location, environmental factors, and nest occupancy on thermoregulatory and ventilation properties.
- Comparing the ventilation processes between mounds with open and closed ventilation system.
- Consolidating insights on thermoregulation and ventilation from various termite species to establish a comprehensive understanding in a cohesive manner.
- Investigating the influence of mound humidity on thermoregulation and ventilation dynamics.

Addressing these research gaps will provide a more comprehensive understanding of termite mounds particularly in relation to thermoregulation and ventilation function.

To advance our understanding of temperature regulation and gas exchange in termite nests, it is important to adopt an interdisciplinary approach where skills from biology, physics, X-ray tomography, engineering and machine learning are combined. X-ray tomography and analysis can help visualise and analyse the structure of mounds at different scales. X-ray tomography can be combined with multiscale numerical simulations to obtain pressure fields, velocity fields, permeability, thermal conductivity, diffusivity, and molecular diffusivity. Furthermore, comprehensive laboratory analyses can be conducted on the acquired mound samples to gain insights into mound behaviour under controlled temperatures, gas concentrations, humidity, and drainage conditions. The flow properties obtained from the mounds can be compared with their structural properties to identify any correlation. Machine learning techniques can also be applied to the mound images, where feasible, to identify patterns and similarities in the arrangement of channels and solid matrices within the mounds, and their relationship with mound properties. This multifaceted approach will improve our insight into termite mound temperature regulation and gas exchange.

## X. Conclusions

The following conclusions can be drawn from this study:

(1) Termite mounds possess the remarkable ability to create a stable microclimate, effectively regulating temperature, gases, and humidity. This regulation is achieved through a combination of passive and active strategies, enabling the creation of an environment conducive to the survival of termites and any associated fungi.

(2) Inhabited mounds exhibit temperature profiles similar to uninhabited mounds, but with slightly elevated temperatures attributed to the metabolic activities of termites and fungi. The contribution of metabolic heat becomes more pronounced in larger mounds. Consequently, it can be inferred that temperature regulation in termite nests is not solely reliant on the termites themselves, but rather on the collective influence of the mound and surrounding soil.

(3) Solar irradiation and the diurnal cycle play crucial roles in shaping the temperature and airflow dynamics within the mound. However, away from the surface of the mound, the correlation between ambient temperature and mound temperature diminishes. These temperature fluctuations in the mound as well as gas transport can be modelled using the heat equation and Fick's law, respectively.

(4) The mound exhibits a trade-off between insulation and effective ventilation. The preference for one over the other depends on various factors including nest location, environmental conditions, and the inhabitants of the mound.

(5) The mounds exhibit micropores of varying sizes in walls, which serve diverse functions such as thermal insulation, $CO_2$ diffusivity, permeability, drainage, and contributing to structural stability.

(6) Using an interdisciplinary approach of combining skills from biology, physics, X-ray tomography, engineering and machine learning can give us insight into temperature regulation and gas exchange in termite mounds addressing numerous research gaps.

## XI. Acknowledgments

N.F.K. and K.S. acknowledge James Watt Scholarship from Heriot-Watt University. N.F.K. also acknowledges Mary Burton funds from Heriot-Watt University. G.T. acknowledges Agence Nationale de la Recherche (ANR-06-BYOS-0008) for funding x-ray medical scanning. G.T. also gratefully acknowledges the Indian Institute of Science to serve as Infosys visiting professor at the Centre for Ecological Sciences in Bengaluru.

Oberst, S., Martin, R., Halkon, B.J., Lai, J.C.S., Evans, T.A. and Saadatfar, M. (2021). Submillimetre mechanistic designs of termite-built structures. *Journal of the Royal Society Interface*, 18: 20200957.

Ocko, S.A., Heyde, A., and Mahadevan, L., 2019. Morphogenesis of termite mounds. *Proceedings of the National Academy of Sciences of the United States of America*, 116 (9): 3379–3384.

Ocko, S.A., King, H., Andreen, D., Bardunias, P., Turner, J.S., Soar, R., and Mahadevan, L., 2017. Solar-powered ventilation of African termite mounds. *Journal of Experimental Biology*, 220 (18): 3260–3269.

Pawlyn, Michael., 2019. *Biomimicry in Architecture*. Routledge.

Perez, R. and Aron, S., 2020. Adaptations to thermal stress in social insects: recent advances and future directions. *Biological Reviews*, 95 (6): 1535–1553.

Perna, A., Jost, C., Couturier, E., Valverde, S., Douady, S., and Theraulaz, G., 2008. The structure of gallery networks in the nests of termite Cubitermes spp. revealed by X-ray tomography. *Naturwissenschaften*, 95 (9): 877–884.

Perna, A. & Theraulaz, G. 2017. When social behaviour is moulded in clay: on growth and form of social insect nests. *Journal of Experimental Biology*, 220: 83-91.

Petrovic AM, Siebert JE & Rieke RE 1982. Soil bulk-density analysis in 3 dimensions by computed tomographic scanning. *Soil Sci Soc Am J*, 46:445–50. available at doi: https://doi.org/10.2136/sssaj1982.03615995004600030001x

Raof, A., Ullah, A., Said, I. Bin, and Ossen, D.R., 2018. Cooling strategies in the biological systems and termite mound: The potential of emulating them to sustainable architecture and bionic engineering, 13(19): 8127-8141.

Räsänen, M., Vesala, R., Rönnholm, P., Arppe, L., Manninen, P., Jylhä, M., Rikkinen, J., Pellikka, P., and Rinne, J., 2023. Assessing CO2 and CH4 fluxes from mounds of African fungus-growing termites. *Biogeosciences Discussions* [preprint]. Available from: https://doi.org/10.5194/bg-2023-24 in review.

Schmidt, A., Jacklyn, P., and Korb, J., 2014. 'Magnetic' termite mounds: Is their unique shape an adaptation to facilitate gas exchange and improve food storage? *Insectes Sociaux*, 61 (1), 41–49.

Schmidt, R.S., 1955. *Termite (Apicotermes) Nests: Important Ethological Material*. Behaviour, 8, (4): 344-356

Singh, K., Muljadi, B.P., Raeini, A.Q., Jost, C., Vandeginste, V., Blunt, M.J., Theraulaz, G. & Degond, P. 2019. The architectural design of smart ventilation and drainage systems in termite nests. *Science Advances*, 5: eaat8520.

T Greves, 1964. Temperature studies of termite colonies in living trees. *Australian Journal of Zoology*, 12: 250–262.

Tapias, A. and Moreno, R. (2020). The effectiveness of computed tomography for the experimental assessment of surfactant-polymer flooding. *Oil & Gas Science and Technology – Revue d'IFP Energies nouvelles*, 75: 5.

Theraulaz, G. and Bonabeau, E., 1999. A brief history of stigmergy. *Artificial Life,* 5: 97–116.

Theraulaz, G., Bonabeau, E. & Deneubourg, J.L. 1998. The origin of nest complexity in social insects, *Complexity*, 3: 15-25

Thomas, R.J. 1981. Ecological studies on the symbiosis of Termitomyces Heim with Nigerian Macrotermitinae. U.K. Univ. London

Turner, Scott. (2000). Architecture and morphogenesis in the mound of Macrotermes michaelseni (Isoptera: Termitidae, Macrotermitinae) in northern Namibia. Cimbebasia. 16. 143-175.